\documentclass[a4paper,11pt]{article}
\usepackage{jinstpub} 
\usepackage{lineno}

\title{\boldmath Monolithic MHz-frame rate digital SiPM-IC with sub-100~ps precision and 70~$\mu$m pixel pitch}

 \author[1]{I. Diehl,\note{Corresponding author.}}
 \author{K. Hansen,}
 \author{T. Vanat,}
 \author[2]{G. Vignola\note{Also at University of Bonn, Germany.},}
 \author{F. Feindt,}
 \author[3]{D. Rastorguev\note{Also at University of Wuppertal, Germany.},}
 \author{and S. Spannagel}

 \affiliation{Deutsches Elektronen-Synchrotron DESY,\\Notkestr. 85, 22607 Hamburg, Germany}
 \emailAdd{inge.diehl@desy.de}

\abstract{This paper presents the design and characterization of a monolithic integrated circuit (IC) including digital silicon photomultipliers (dSiPMs) arranged in a 32$~\times~$32 pixel matrix at 70$~\mu$m pitch. The IC provides per-quadrant time stamping and hit-map readout, and is fabricated in a standard 150-nm CMOS technology. Each dSiPM pixel consists of four single-photon avalanche diodes (SPADs) sharing a quenching and subsequent processing circuitry and has a fill factor of 30$~\%$. A sub-100$~$ps precision, 12-bit time-to-digital converter (TDC) provides timestamps per quadrant with an acquisition rate of 3$~$MHz. Together with the hit map, the total sustained data throughput of the IC amounts to 4$~$Gbps. Measurements obtained in a dark, temperature-stable environment as well as by using a pulsed laser environment show the full dSiPM-IC functionality. The dark-count rate (DCR) as function of the overvoltage and temperature, the TDC resolution, differential and integral nonlinearity (DNL/INL) as well as the propagation-delay variations across the matrix are presented. With aid of additional peripheral test structures, the main building blocks are characterized and key parameters are presented.}


\begin{document}
	\maketitle
	\flushbottom
	\section{Introduction} \label{sec:intro}
	
Silicon photomultipliers (SiPMs) are composed of an array of single-photon avalanche diodes (SPADs), which are photodiodes reverse biased above breakdown voltage, operating in Geiger-mode. SiPMs working principles and main advantages like large intrinsic gain (typically 10$^{5}$ to 10$^{6}$), insensitivity against magnetic fields, and large dynamic range starting with single-photon response at low bias voltages, are well explained and specified in literature~\cite{1621377, Otte2006TheSP,bruschini2019singlephoton}. Arrays of SPADs are highlighted as ideal candidates when high sensitivity is required together with high frame rate and precise timing resolutions~\cite{5698906}. As stated in ~\cite{s22082919}, the capability of photon counting makes SPADs the detector of choice for applications in which conventional photodiodes and charge-coupled devices cannot be used, and a large number of applications for SPAD arrays are listed. Beside the usage as imaging device, e.g. in~\cite{morimoto2019megapixel}, also light detection and ranging~\cite{8470112}, and direct minimum ionizing particle detection~\cite{gramuglia2021sub10} are mentioned.

In traditional analog SiPMs, each single SPAD is connected to its quenching resistor forming a so-called microcell. All microcells are connected in parallel and the common current output signal has to be amplified and digitized. The number of microcells defines the dynamic range. In digital SiPMs (dSiPMs) one benefits from the inherent digital behavior of the device, when sensing directly the output of an individual SPAD by an embedded quenching and recharging circuitry together with a simple discriminator, e.g. an inverter. This concept was initially introduced in~\cite{5402143}, where SPADs are integrated with conventional CMOS circuits on the same substrate. This additional circuitry, in pixel or at the periphery can be used to acquire, store and transmit data. It negatively impacts the fill factor~\cite{bruschini2019singlephoton}, but offers the information which SPAD is hit. Individual readout permits the definition of spatial granularity, e.g. grouping of SPADs in pixel, reduces circuit complexity and increases the fill factor. Furthermore, it offers temporal granularity by delivering the timestamp for a section of the pixel array, and it provides the opportunity to identify and switch off noisy pixels, to adapt the hold-off time minimizing the afterpulsing probability, and to count hits within a pixel. The main drawback of implementing SPADs in standard CMOS technologies compared to a custom technology is the relatively high dark noise, expressed in terms of dark-count rate (DCR)~\cite{TORILLA2023167693}. To compensate these DCR and fill-factor issues for the detection of charged particles, one could implement a dual layer structure of different SPAD arrays for a coincidence measurement, like in~\cite{Ratti}. DCR is one of the most important properties of SPADs, besides photon-detection probability (PDP), representing the avalanche probability of the device in response to a photon absorption at a given wavelength, and afterpulsing, which introduces false events that are correlated in time to previous detection~\cite{bruschini2019singlephoton}. Also optical crosstalk, fill factor, timing resolution and deadtime have to be considered to characterize SPADs.

The presented dSiPM-IC uses a similar readout concept like in the preceding project published in~\cite{7581816,8824395}. The concept comprises a 32~$\times$~32~dSiPM-pixel matrix subdivided into quadrants. Each quadrant consists of a 16~$\times$~16~dSiPM-pixel array sharing a time-to-digital converter (TDC) and a validation logic. The 12-bit TDC is designed to provide the timestamp of the fastest pixel with a time resolution of less than 100~ps. The validation logic discards undesirable events by setting a threshold. For example, if only one pixel is hit at the same time, it will be most likely a dark event. Along with the timing information, also the hit map can be read out continuously in a frame-based mode at 3~MHz. The previous readout chip was designed in Global Foundry's 130-nm CMOS technology comprising a single quadrant. The major difference between the old and new concept is the interconnection approach of sensor and readout electronics. In the old concept, the readout IC is flip-chip connected to the sensor chip utilizing 30~$\mu$m solder spheres at 50~$\mu$m pitch~\cite{FCbonding}. This hybrid approach enables the direct readout of each sensor pixel by its corresponding pixel electronics occupying the same pixel area at about 50~$\mu$m pitch, without affecting the fill factor. Because the sensitive area of the sensor chip is now vis-á-vis with the readout chip, direct light illumination of the matrix is not possible. In~\cite{8824395}, DCR measurements were carried out on the first hybrid samples, and the TDC characterized. In concluding, it was planed to reduce the high DCR by a new sensor design. A further required step was the redesign of the IC in another CMOS process. It was decided to follow the current trend and select a monolithic approach~\cite{sonneveld2023design}. The chosen process provides on the one hand fully characterized SPADs in four configurations with suitable properties, and on the other hand the possibility to redesign the previous developed readout electronics in a comparable process. The advantage of having sensor and readout electronics on the same device is the designer's independency of sensor provider and interconnection issues. This allowed a fast and cost-efficient realization of a first proof-of-principle device.
\newpage
The realized IC is shown and its essential components are described in section~\ref{sec:design}. Section~\ref{sec:measure} summarizes the results obtained from measurements and compares results from test-circuit blocks with results from earlier prototypes~\cite{8824395}. Furthermore, the used Caribou data acquisition (DAQ) system~\cite{Vanat:2020nS} and the test setups are introduced. The conclusions are given in section~\ref{sec:conclude}.

	\section{Chip architecture} \label{sec:design}
\subsection{Matrix and periphery}
\begin{figure}[h]
	\centering
	\includegraphics[width=0.93\linewidth]{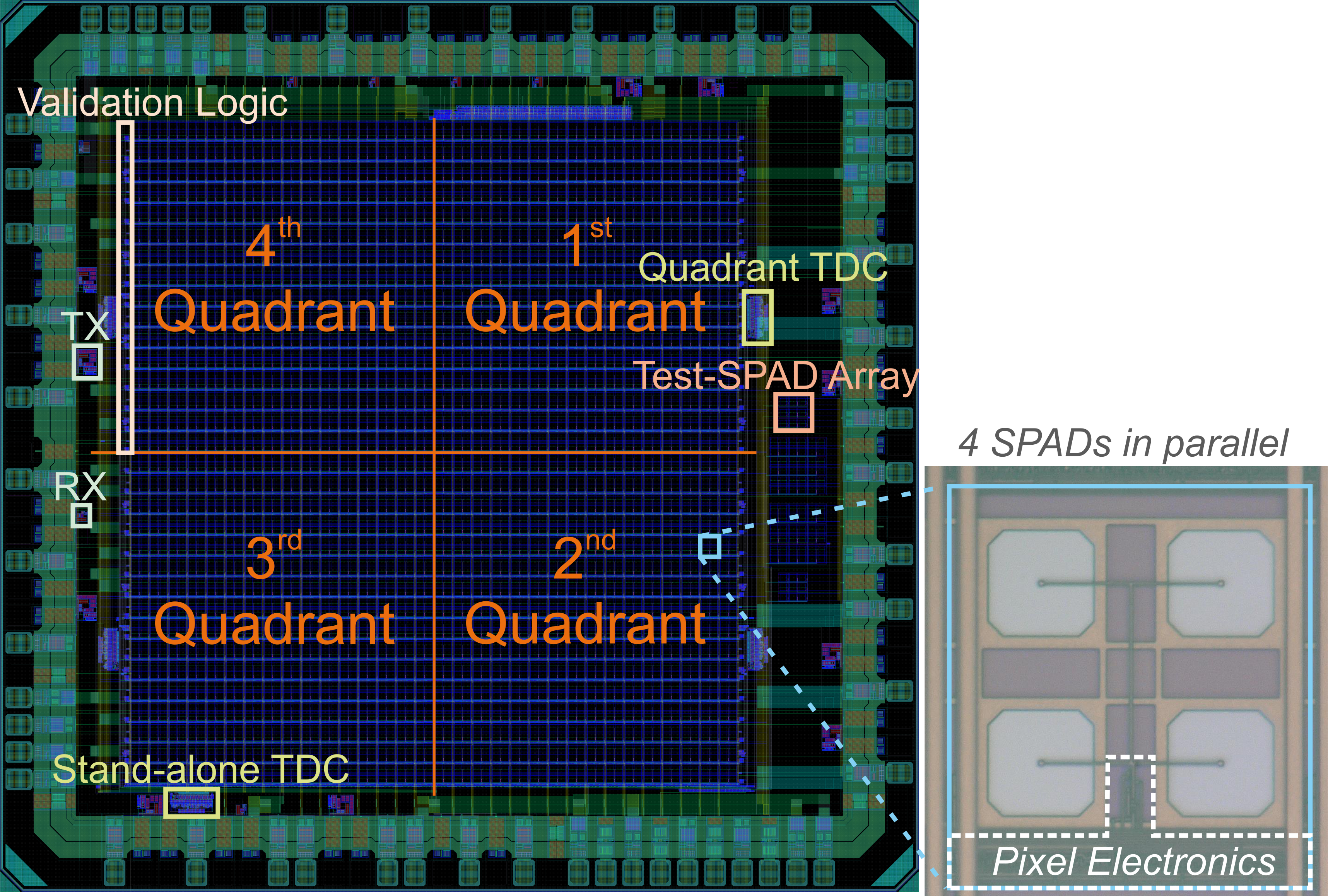}
	\put(-410,264){\makebox(0,0){a)}} 	\put(-108,154){\makebox(0,0){b)}}
	\caption{a) IC layout (3400~$\times$~3300~$\mu$m$^{2}$). b) dSiPM-pixel photograph (69.6~$\times$~76~$\mu$m$^{2}$).}
	\label{fig:chip}
\end{figure}

The monolithic IC (cf. figure~\ref{fig:chip}a) has been designed in LFoundry's 150-nm CMOS technology to take advantage of an available add-on library including single p+/nwell SPADs with an active area of 20~$\times$~20~$\mu$m$^{2}$. This area is framed by a cathode ring plus an additional pwell ring to minimize the crosstalk to neighboring cells. The dSiPM-pixel design comprises four SPADs in parallel sharing the pwell ring and using it partially for the NMOS transistors of the common pixel electronics to reduce dead area. In this way, the pixel covers an area of 69.6~$\times$~76~$\mu$m$^{2}$ with a fill factor of 30~$\%$. Figure~\ref{fig:chip} shows the layout of the IC (a) and a photograph of the dSiPM pixel (b).

\begin{figure}[htb]
	\centering
	\includegraphics[width=1\linewidth]{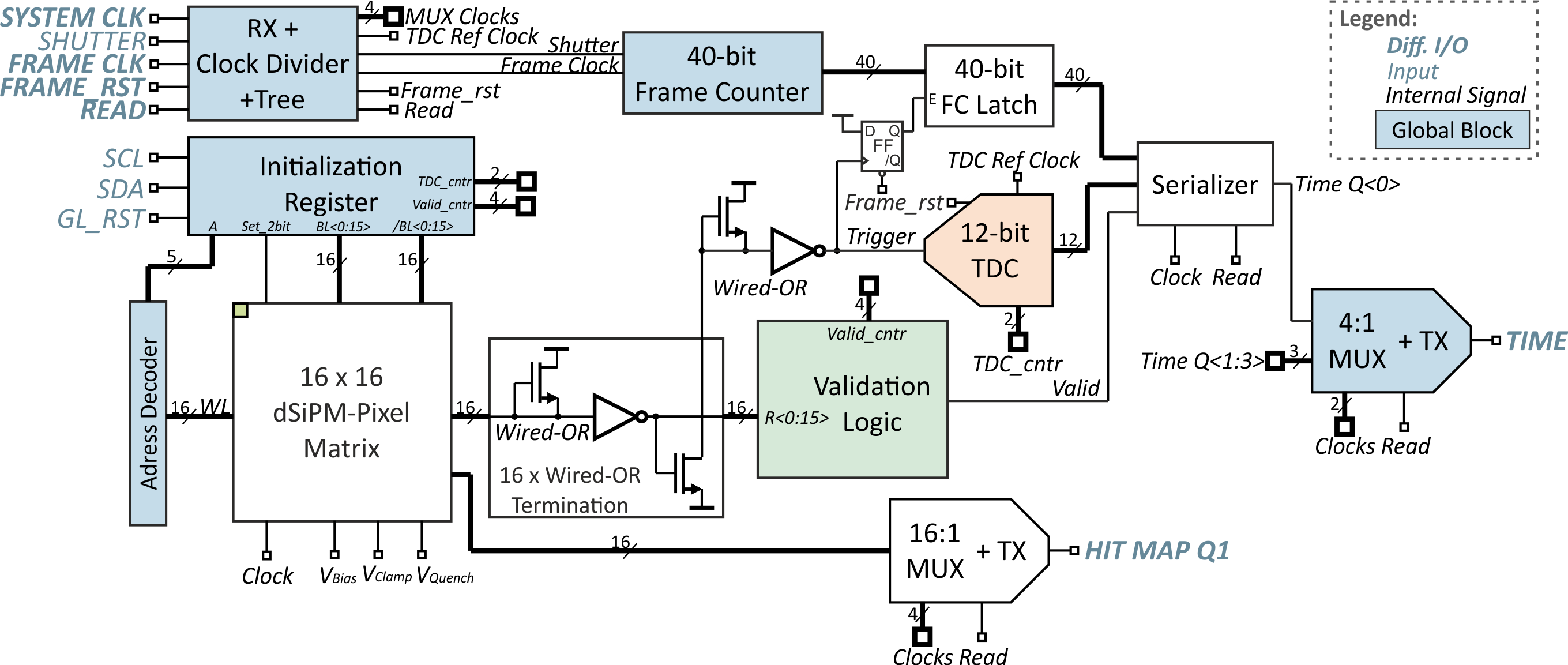}
	\caption{Quadrant block diagram.}
	\label{fig:ROscheme}
\end{figure}

The IC consists of four identical units (quadrants), each one (cf. figure~\ref{fig:ROscheme}) with a 16~$\times$~16 dSiPM-pixel matrix, a single 12-bit TDC for event-time stamping, a validation logic with adjustable settings for discarding undesirable events, and serializer circuits followed by links (MUX~+~TX) for fast (about 1~Gbps) and sustained data readout. In this way, the system provides the full hit map of a 32~$\times$~32~dSiPM-pixel matrix together with four timestamps for each frame.
		
Figure~\ref{fig:ROscheme} shows a simplified block diagram of one quadrant with all components, as well as additional blocks used globally for all quadrants. 
When an avalanche process is initiated in a SPAD, the pixel draws current from a fast wired-OR connection, and the earliest pixel in the readout frame triggers the TDC. Simultaneously, the row-wise wired-ORs are monitored by a validation logic generating a valid bit for this event. The latched outputs of a peripheral 40-bit frame counter (FC) complete the quadrant timing information with the frame number of the event. The TDC and FC are started by the \emph{Shutter} signal defining the beginning of a measurement. The 40-bit FC allows a total recording time of about 100~hours for a measurement. The timing data are serialized and multiplexed together with the timing data of the other quadrants. The counted hits in each pixel are serialized row-wise and multiplexed column-wise to deliver the hit map of the quadrant. 
In the periphery, an initialization register provides the control signals for pixel masking, validation thresholds and TDC settings. Furthermore, clock dividers and buffers serve to distribute all essential clocks to the components. The main clocks are provided by the DAQ system. The frame-based readout and operation mode runs at a targeted frequency of 3~MHz (\emph{FRAME CLK}). Additionally, a 408-MHz \emph{SYSTEM CLK} (multiple of the \emph{FRAME CLK}) is required as reference clock for the TDCs (cf. \emph{TDC Ref Clock}) as well as for the multiplexers to enable the sustained readout of the entire hit matrix.

\subsection{Block details and timing}

\begin{figure}
	\centering
	\includegraphics[width=.9\linewidth]{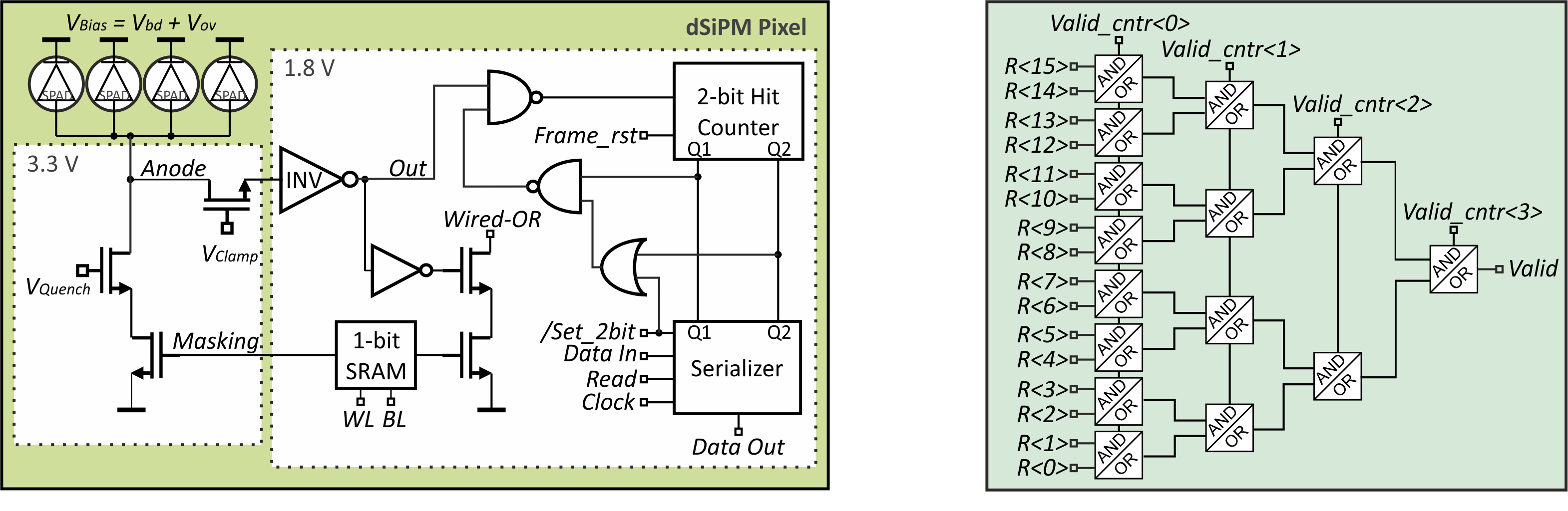}
	\put(-400,120){\makebox(0,0){a)}} 	\put(-155,120){\makebox(0,0){c)}}
	\qquad
	\centering
	\includegraphics[width=1\linewidth]{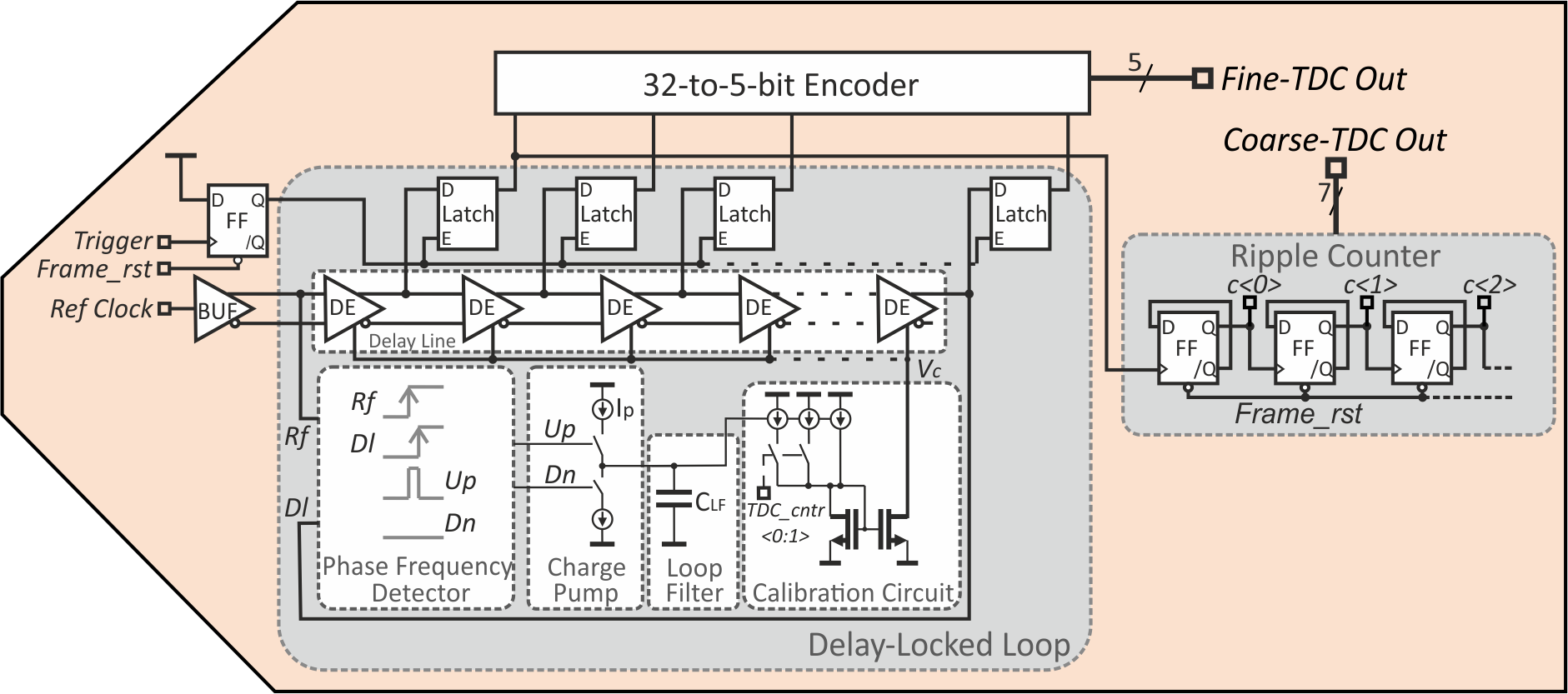}
	\put(-420,170){\makebox(0,0){b)}}
	\qquad
	\caption{Simplified equivalent circuits of: a) the dSiPM pixel, b) the TDC, and c) the validation logic.}
	\label{fig:circuits}
\end{figure}
	
Figure~\ref{fig:circuits}a shows the dSiPM-pixel electronics consisting of the four SPADs connected in parallel, a front-end circuitry using 3.3-V NMOS transistors, and a readout circuitry operating at core voltage of 1.8~V. The front end allows an overvoltage (V$_{ov}$) of maximally 3.6~V, which sets the anode voltage. The used quenching circuit is similar to the presented one in~\cite{7066818}, where the quenching is performed by a globally biased transistor (cf. \emph{V$_{Quench}$}) and an inverter (cf. INV) as comparator for the digital pulse shaping (cf. \emph{Out}). A clamping transistor (cf. \emph{V$_{Clamp}$}) limits the inverter input to 1.8~V. The pixel electronics include the possibility to mask a pixel (cf. \emph{Masking}) via a single SRAM cell. The hold-off time of the pixel can be adapted by the global bias voltage \emph{V$_{Quench}$}. Detected hits can be counted within the acquisition window by the 2-bit hit counter. If this function is not needed, the counter can be set as buffer by the control signal \emph{/Set$\_$2bit}.
	
Figure~\ref{fig:circuits}b shows the TDC, which is divided into a fine and coarse converter. The fine TDC consists of a delay-locked loop (DLL) with 32 differential delay elements (DEs). A phase-frequency detector compares the incoming reference signal (cf. \emph{Rf}) with the delayed one (cf. \emph{Dl}) and steers the following charge pump like in~\cite{DLL}. The resulting voltage over the loop-filter capacitance (C$_{LF}$) together with a calibration circuit controls the delay of all DEs (cf. \emph{V$_{c}$}). The DE outputs are latched by the wired-OR trigger signal until the next frame starts. The latched signals are encoded as thermometer code and have to be decoded into binary with the aid of a 32-to-5-bit encoder. After each DLL cycle, a subsequent ripple counter is incremented and serves as 7-bit coarse TDC.
	
The row-wise wired-OR outputs (cf. \emph{R<0:15>}) are monitored by a 4-step validation logic generating a valid bit for an event, as displayed in figure~\ref{fig:circuits}c. In the first validation step, every row output is connected by a selectable AND/OR gate with its neighboring row output. In each subsequent step all gates in a column can be programmed as AND or OR (by \emph{Valid$\_$cntr<0:3>}). In this way, pixel in a cluster of few pixels firing simultaneously can be identified. 
		
\begin{figure}[h]
	\centering
	\includegraphics[width=1\linewidth]{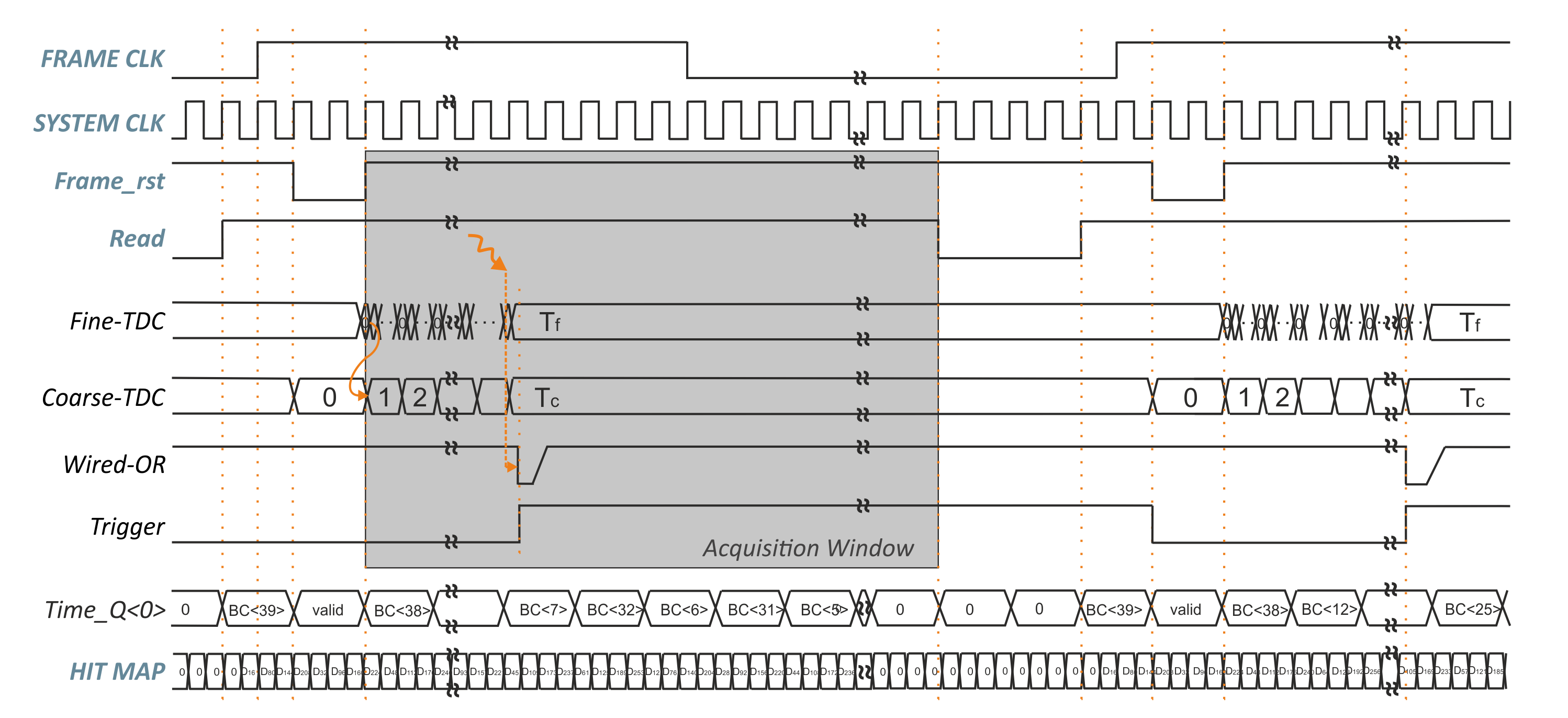}
	\caption{Timing diagram of the frame-based readout.}
	\label{fig:Timing}
\end{figure}

Figure~\ref{fig:Timing} shows the underlying timing diagram with input clocks, internal signals and output data for the hit counter set in 1-bit mode. The acquisition window is defined by the rising edge of \emph{Frame$\_$rst} and the falling edge of \emph{Read}. Afterwards, all collected data are buffered into the serializers and read out during the following frame. The dynamic range of the TDC with $2^{7}$/408~MHz~=~313.7~ns corresponds to the acquisition window. By using the hit counter in 2-bit mode, the readout time has to be adapted to enable full hit-map readout. Accordingly, the acquisition window extends to two frame-clock cycles.
 	
\subsection{Test circuits}
 	
\begin{figure}[h]
	\centering
	\includegraphics[width=1\linewidth]{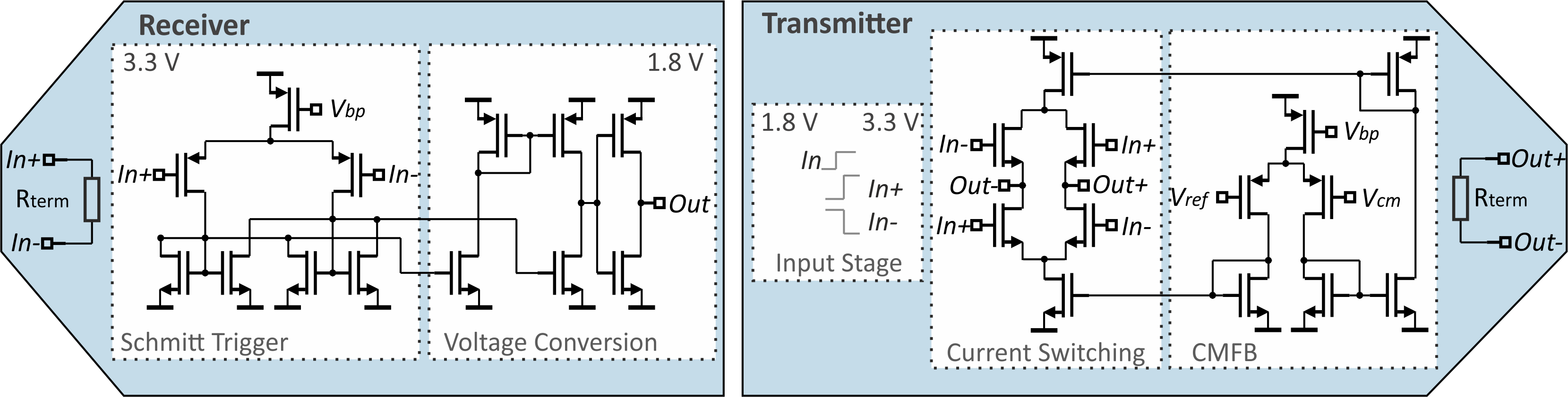}
	\caption{Simplified block diagrams of the RX and TX.}
	\label{fig:RXTX}
\end{figure}

In the periphery of the matrix, test circuits are integrated for single-block characterization (cf. figure~\ref{fig:chip}a). These extra blocks comprise a stand-alone TDC, as well as a chain of receiver~(RX) and transmitter~(TX) links using the low-voltage differential signaling~(LVDS) standard. The simplified block schemes are shown in figure~\ref{fig:RXTX}. The RX consists of a Schmitt-Trigger circuitry operated at 3.3~V supply voltage and is followed by a voltage-conversation stage delivering single-ended output. The TX includes an input stage comprising an edge aligner, voltage converter and buffers to drive the input of the subsequent current-switching circuitry. To keep the output common mode stable about 1.2~V, a common-mode feedback (CMFB) steers the output current. Both circuits are terminated with 100~$\Omega$ on the chip.
 		
Additionally, sensor-test structures in different configurations are implemented. For example, a 3~$\times$~3~test-SPAD array, where the center SPAD is connected to the same front end as used for the dSiPM pixel, enables the monitoring of the digital output of a single SPAD.
	\section{Measurements} \label{sec:measure}
	
\subsection{Experimental setup}	

For characterization of the IC, the versatile Caribou readout system is used enabling fast and low-cost implementation of new solid-state detector prototypes. It offers open access to hardware, firmware~\cite{caribou} and software~\cite{peary} speeding up the test setup development. Caribou mainly consists of a system-on-chip (SoC) evaluation board (Xilinx~ZC706) and a control-and-readout (CaR) interface board. A field-programmable gate array (FPGA) runs custom hardware blocks for data processing and detector controlling, and an embedded CPU runs the DAQ and control software. The CaR board provides the physical interface between the SoC and the detector. It also includes power supplies, analogue and digital I/Os, a clock generator, analog-to-digital converters, current and voltage references as well as several connectors. A detector-specific carrier board comprises the IC, LVDS repeater, and components for filtering, decoupling and line termination, as well as a trigger logic for an external pulsed laser source. The carrier board is covered by an aluminum case protecting the dSiPM-IC from physical damage, external light and acting as a heat sink.
The dark-event based characterization of the dSiPM-pixel matrix was performed by using a climate chamber and temperature sets between $-$25$^{\circ}$C and 25$^{\circ}$C. 
	
In order to determine the propagation delay along the wired-OR lines, a pulsed 1054-nm laser was operated synchronously to the frame clock. The laser is movable in all dimensions illuminating sequentially the actual enabled pixel with a spot diameter of $\sim$0.5~mm.
	
Another IC test board together with the data-timing generator DTG~5334 (3.35~GHz clock, 0.2~ps step size) and the oscilloscope Teledyne LeCroy SDA-760ZI were used as second test setup for the characterization of the separate links and the stand-alone TDC at room temperature.

\subsection{Results}	
\subsubsection{dSiPM-pixel matrix}
	
With the Caribou-test setup, several samples were measured. Initially, the DCR was monitored as a function of the common SPAD-bias voltage (V$_{bias}$) by accumulating dark hits within 10,000~frames. A hit can only be detected by the dSiPM pixel (cf. figure~\ref{fig:circuits}a) when V$_{bias}$ exceeds the breakdown voltage (V$_{bd}$) plus the threshold voltage of the inverter. Taking this into account, V$_{bd}$ could be estimated by taking the voltage at which hits are detected minus the threshold voltage of the inverter, which is about 0.6~V. The voltage above V$_{bd}$ is defined as the overvoltage (V$_{ov}$). 
\begin{figure}[h]
	\centering
	\includegraphics[width=.47\linewidth,trim=5 0 30 20,clip]{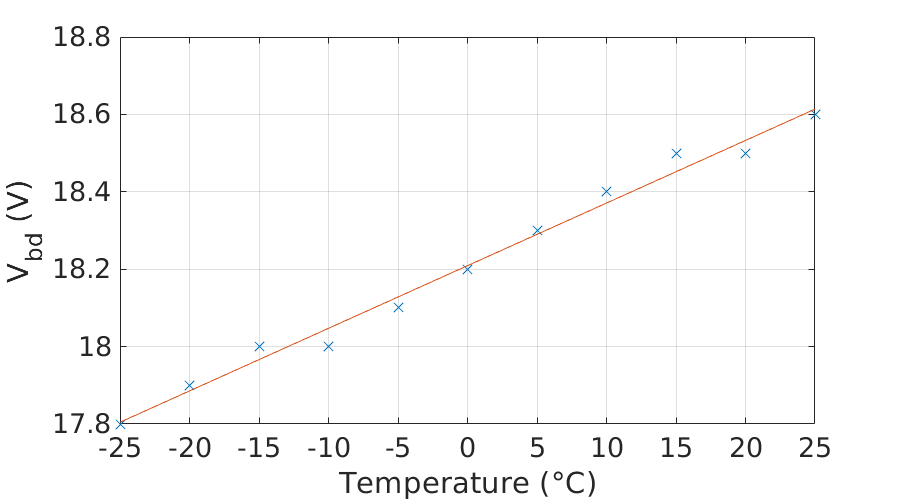}
	\put(-200,107){\makebox(0,0){a)}}
	\qquad
	\includegraphics[width=.47\linewidth,trim=25 5 35 0,clip]{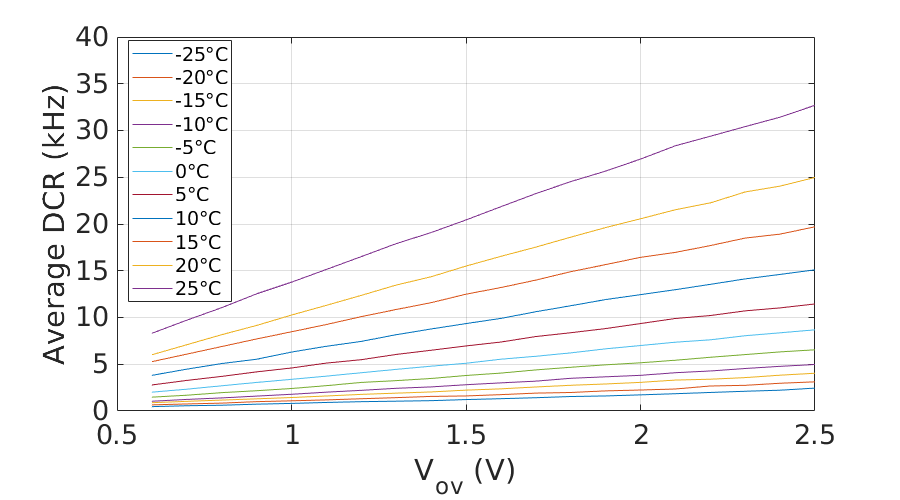}
	\put(-202,107){\makebox(0,0){b)}}
	\qquad
	\includegraphics[width=.47\linewidth,trim=8 0 35 10,clip]{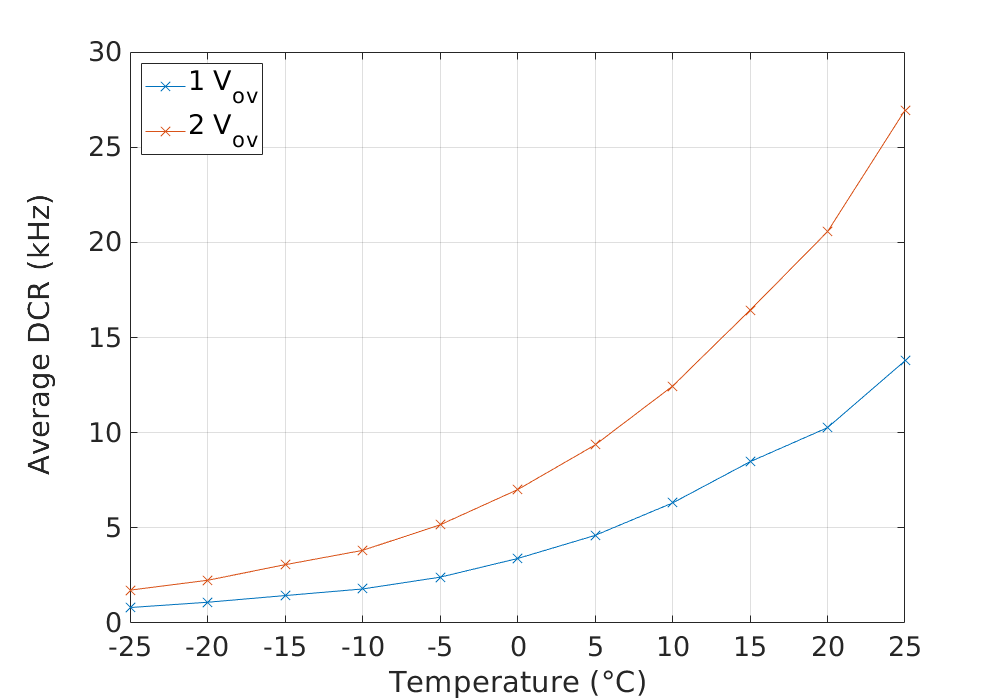}
	\put(-200,137){\makebox(0,0){c)}}
	\color{red}	\put(-177,37){\vector(0,-1){12}}	\put(-93,58){\vector(0,-1){12}}	\put(-10,142){\vector(0,-1){12}}
	\color{black}
	\qquad
	\includegraphics[width=.47\linewidth,trim=25 0 40 10,clip]{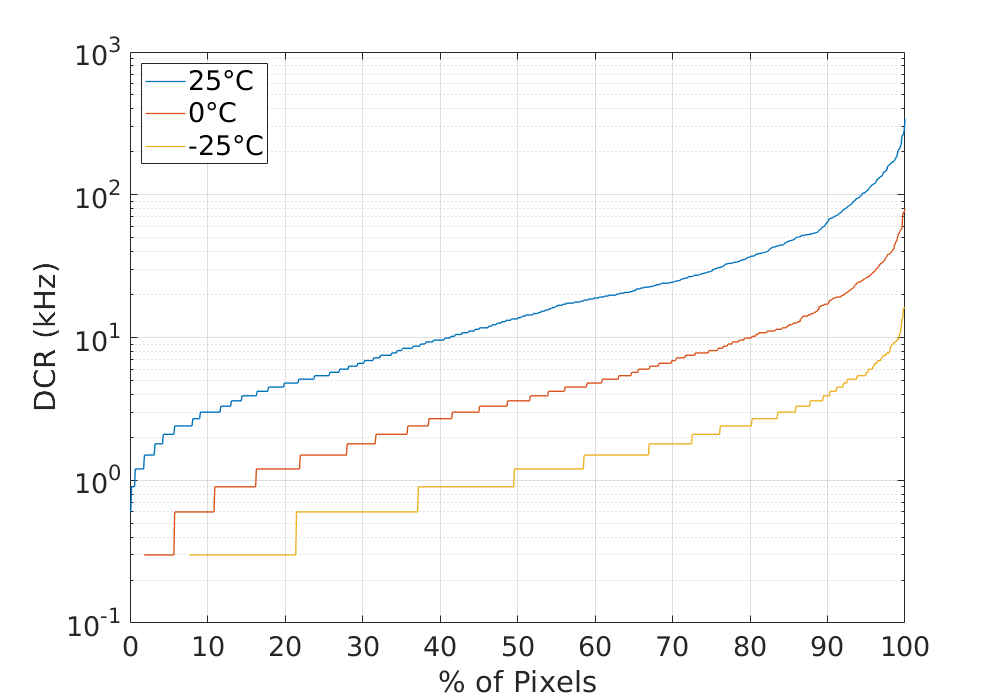}
	\put(-202,137){\makebox(0,0){d)}}
	\color{red}	\put(-65,69){\vector(0,-1){12}}
	\put(-59,88){\vector(0,-1){12}}	\put(-56,106){\vector(0,-1){12}}
	\color{black}
	\caption{a) V$_{bd}$ versus temperature. Average DCR of all pixels: b) versus V$_{ov}$ for ten temperature settings, and c) versus temperature at V$_{ov}$~=~1~V and 2~V. d) Cumulative distribution of all pixel-DCRs in the matrix at three temperatures and V$_{ov}$~=~2~V.}
	\label{fig:dcr}
\end{figure}
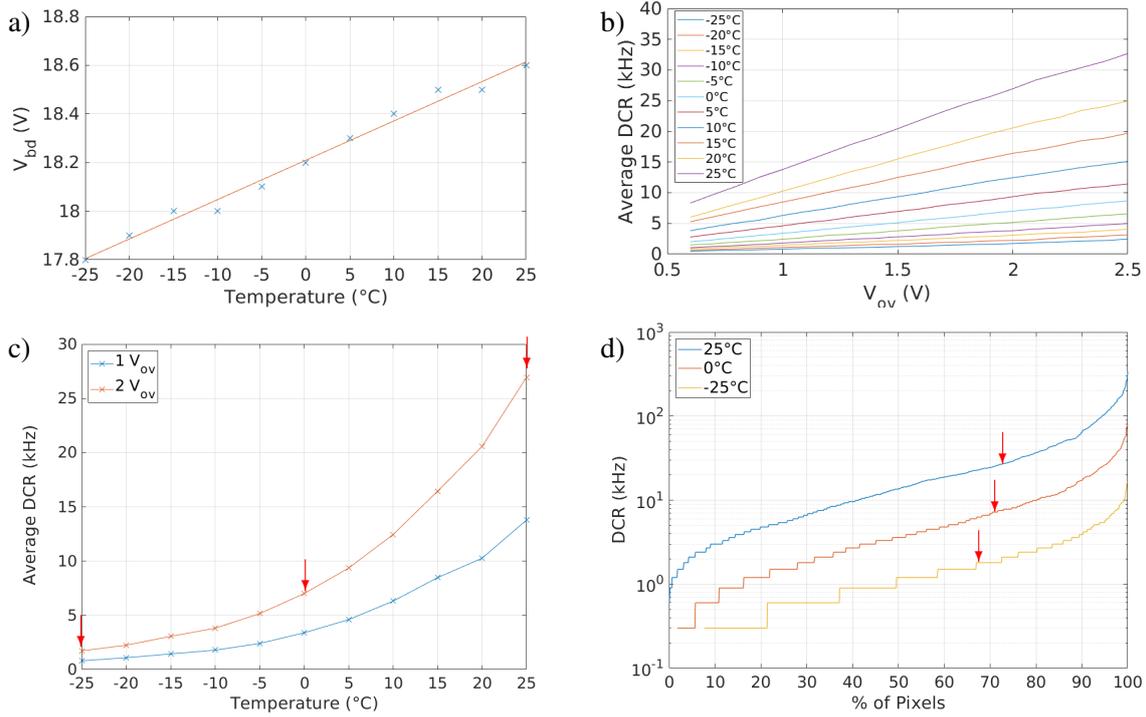
	
All results of a representative sample are shown in figure~\ref{fig:dcr}. V$_{bd}$ as a function of temperature is plotted in figure~\ref{fig:dcr}a and the average DCR of all 1024 pixels is plotted in~figure~\ref{fig:dcr}b. This diagram illustrates the strong dependency of DCR on temperature and V$_{ov}$. For example, a decrease of V$_{ov}$ from 2~V to 1~V results in a DCR reduction of about 50~$\%$. Further cooling of the sample reduces the noise behavior enormously. This is nicely seen in figure~\ref{fig:dcr}c for two different overvoltages. While in~\ref{fig:dcr}c the average DCR of all pixels within the matrix is plotted vs. temperature, figure~\ref{fig:dcr}d shows the pixel-DCRs for $-$25$^{\circ}$C, 0$^{\circ}$C and 25$^{\circ}$C as cumulative distribution (100~$\%$: full matrix of 1024~pixels) for V$_{ov}$~=~2~V.
This plot illustrates the wide spread over the whole matrix, ranging from 600~Hz up to 339.3~kHz for 25$^{\circ}$C and 200~Hz up to 16.2~kHz for $-$25$^{\circ}$C. As a comparison, in \cite{9142240} a very similar distribution for an array of 32~$\times$~32 SPADs at 44.64~$\mu$m pitch using the same sensor cell and process is shown. The range is here between some ten~Hz and $\sim$200~kHz. By using our peripheral 3~$\times$~3 test-SPAD array we were able to perform the DCR measurements on single SPADs. The results of three SPADs (550~Hz, 1050~Hz and 15.3~kHz) fit into this distribution plot.
If we take the data from figure~\ref{fig:dcr}c (tagged with a "$\color{red}\downarrow$") and look for these in figure~\ref{fig:dcr}d, one can say that at least 67~$\%$ of the pixels at $-$25$^{\circ}$C and 73~$\%$ of the pixels at 25$^{\circ}$C have DCR values below the average ones. This percentage means that only a few very noisy pixels boost the average DCR value, and this behavior grows with temperature. Disabling 10~$\%$ of the noisiest pixels leads to an improvement of 40~$\%$ in average DCR. 
	
\subsubsection{dSiPM-pixel electronics}

\begin{figure}[t]
	\centering
	\includegraphics[width=0.8\linewidth,trim=20 0 20 30,clip]{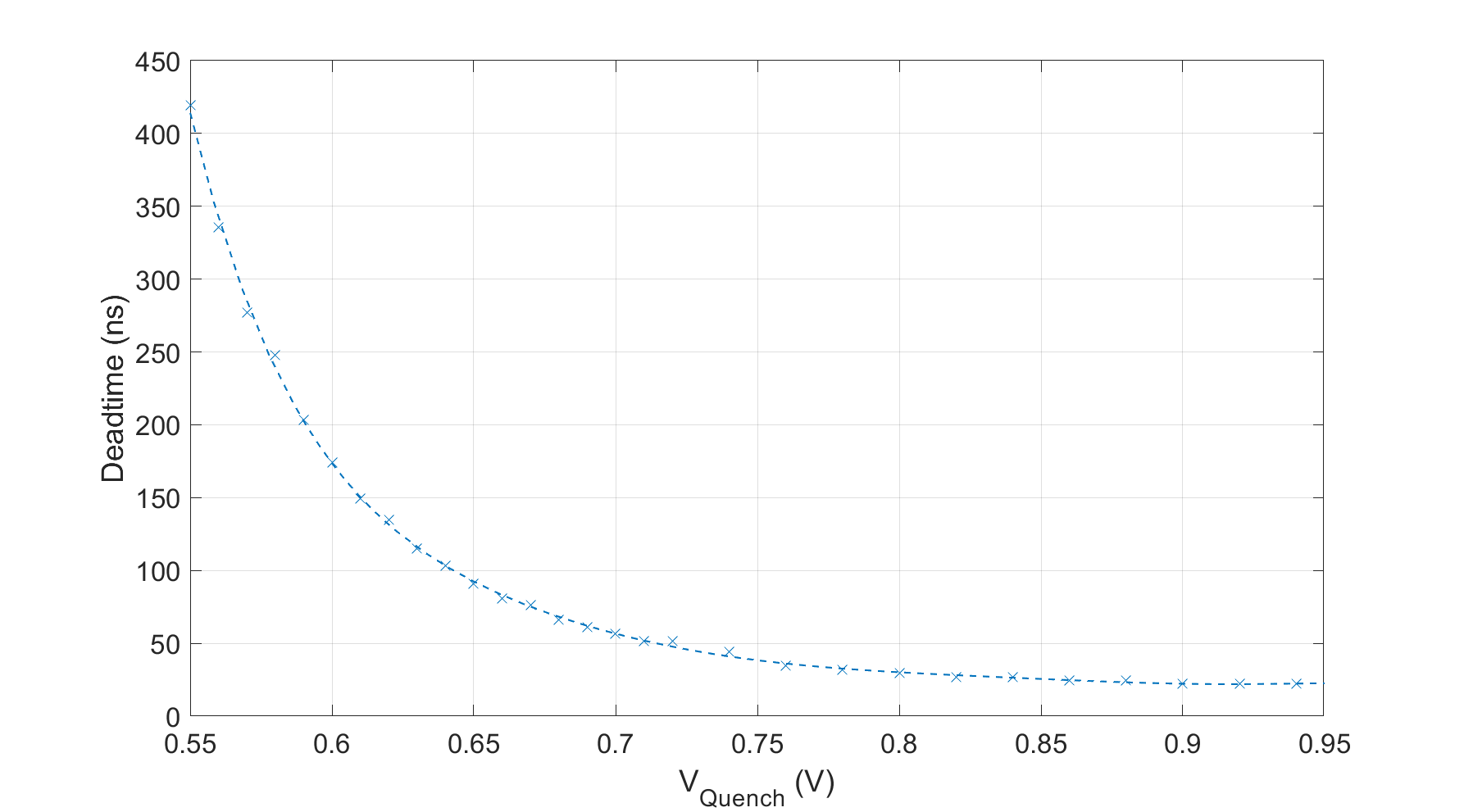}
	\caption{Deadtime versus V$_{Quench}$, with V$_{ov}$~=~2~V at room temperature.}
	\label{fig:dead}
\end{figure}

The previous measurements were taken without using the 2-bit hit-counting functionality in pixel. This feature allows for the determination of the deadtime of the pixel (quenching and recharging). 
Figure~\ref{fig:dead} shows the deadtime as function of the global bias voltage (V$_{Quench}$) of the quenching transistor (cf. figure~\ref{fig:circuits}a), exemplary for a pixel in the center of the laser spot. For this measurement, two laser pulses were send within a frame starting with maximal distance (about 450~ns) to each other and minimal V$_{Quench}$. Below V$_{Quench}$~=~0.55~V, the electronics counts only one hit, that means the pixel needs more time for recharging than the acquisition window allows. After first recognition of the second hit, the deadtime can be defined as distance between both input pulses. This distance can be decreased with increasing V$_{Quench}$. The minimal deadtime is achieved at V$_{Quench}$~=~0.95~V with about 22~ns.
	
\subsubsection{Quadrant TDC}
	
\begin{figure}[b]
	\centering
	\includegraphics[width=0.98\linewidth]{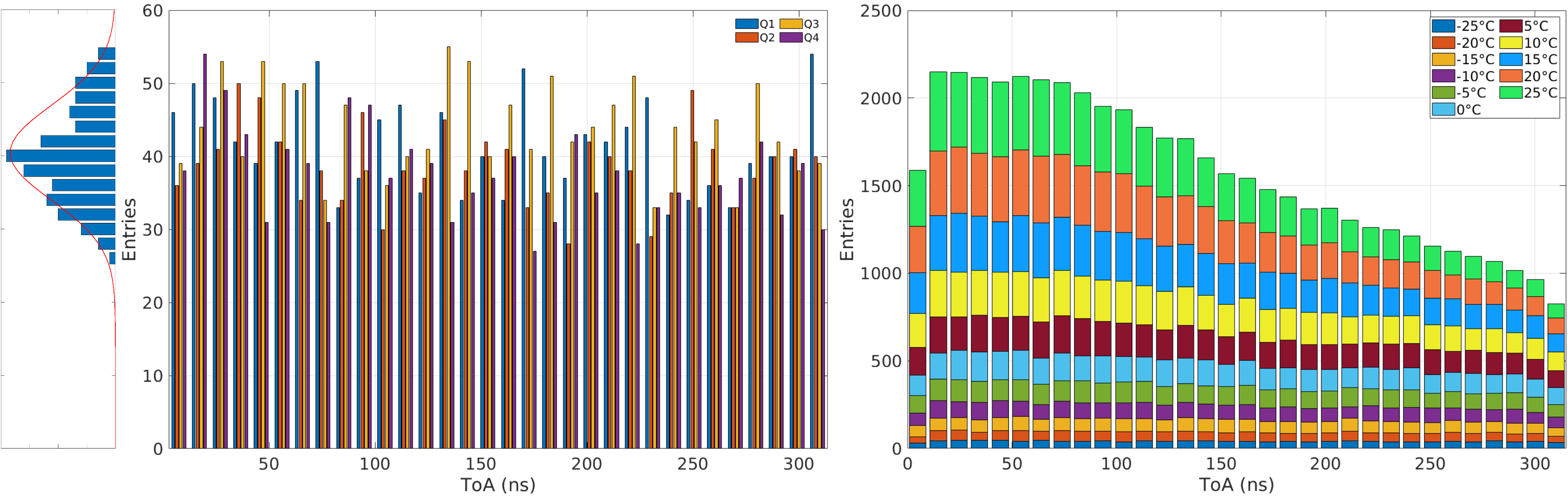}
	\put(-426,125){\makebox(0,0){a)}}
	\put(-194,125){\makebox(0,0){b)}}
	\caption{a) Histograms of all four quadrant-TDC data of the dark-event measurements at V$_{ov}$~=~2~V and $-$25$^{\circ}$C, and their Gaussian fit. b) Collected histograms of all TDC data for $-$25$^{\circ}$ to 25$^{\circ}$C.}
	\label{fig:ToA}
\end{figure}

In addition to the hit map, the individual quadrant-TDC data are monitored during the dark-event measurements. In figure~\ref{fig:ToA}, time-of-arrival (ToA) is plotted into histograms measured at V$_{ov}$~=~2~V and different operation temperatures. At $-$25$^{\circ}$C (figure~\ref{fig:ToA}a), the entries for all ToA values are roughly uniformly distributed over the dynamic range (cf. the Gaussian fit on the left of figure~\ref{fig:ToA}a). With increasing temperature, the entries shift to lower timestamp values (cf. figure~\ref{fig:ToA}b). This is mainly caused by the temperature-dependent DCR. The probability for more entries at lower timestamps increases with number of firing pixels within a quadrant, because all of them share the same TDC, but only the fastest one defines the timestamp.

\begin{figure}[h]
	\centering
	\includegraphics[width=1\linewidth,trim=20 30 20 0,clip]{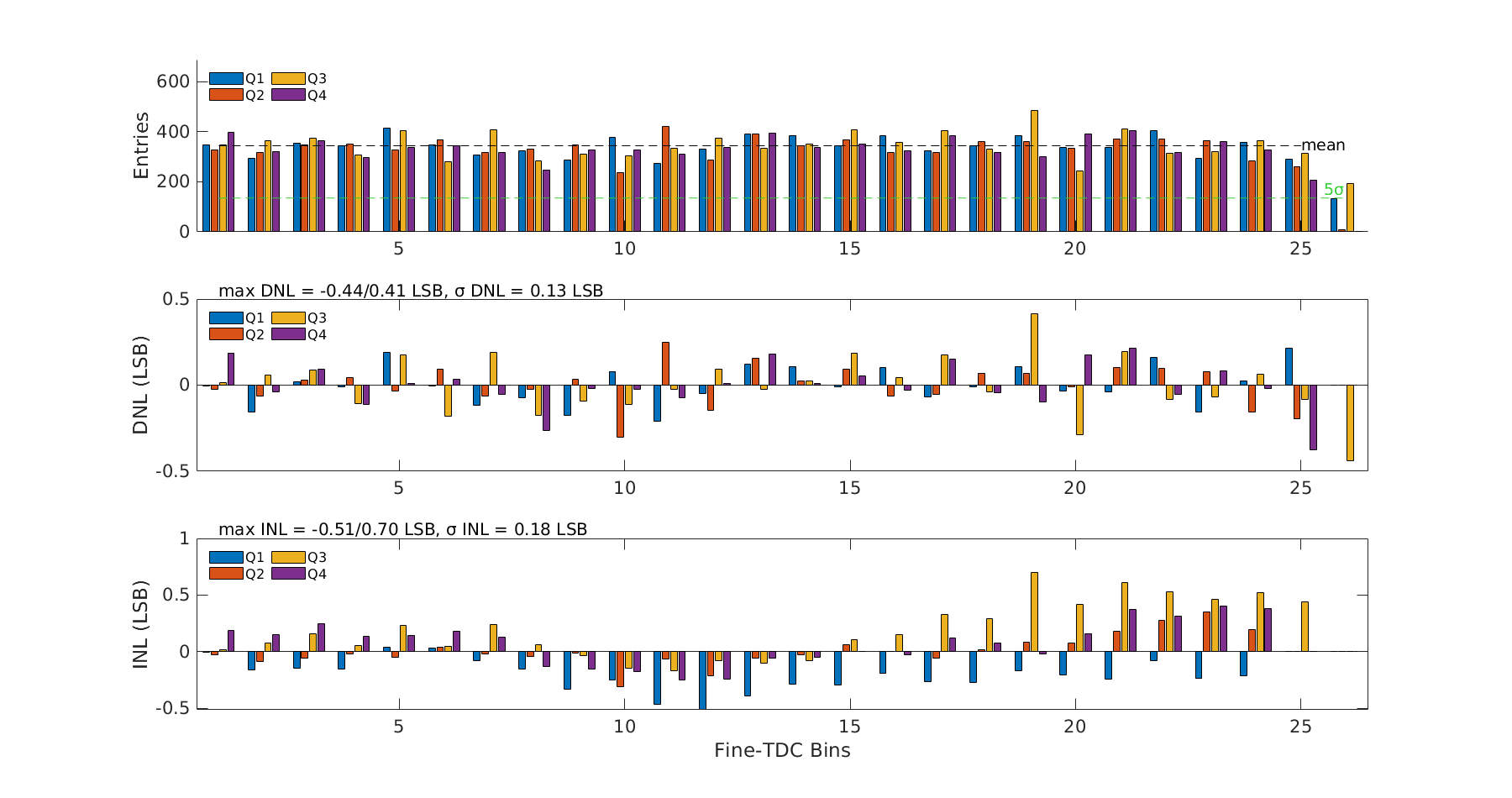}
	\put(-410,200){\makebox(0,0){a)}}
	\put(-410,130){\makebox(0,0){b)}}
	\put(-410,60){\makebox(0,0){c)}}
	\qquad
	\centering
	\includegraphics[width=1\linewidth,trim=20 460 20 40,clip]{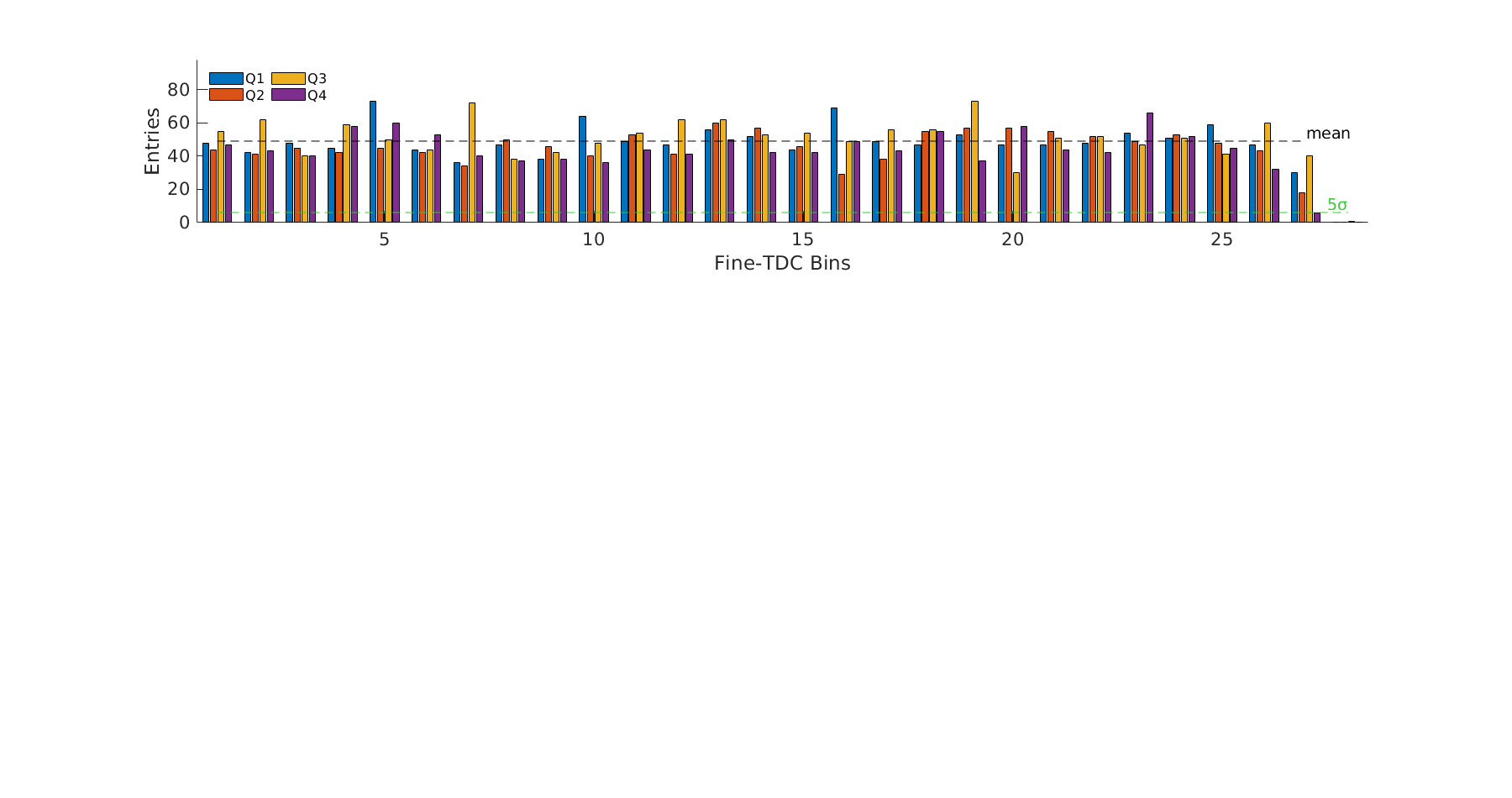}
	\put(-410,60){\makebox(0,0){d)}}
	\caption{a) Histograms, b) DNL, and c) INL of all quadrant fine-TDC data of the dark-event measurements at V$_{ov}$~=~2~V and 25$^{\circ}$C. d) Histograms at $-$25$^{\circ}$C.}
	\label{fig:Hist}
\end{figure}

Due to the non-uniform distribution it is not possible to determine the integral and differential nonlinearity (INL and DNL) of the TDCs over their total dynamic range with aid of the statistical code density test. Therefore, figure~\ref{fig:Hist}a only shows the histograms of the fine-TDC values. The deviation of entries for every bin from the mean value of all entries defines the DNL of each bin. The INL can be calculated as the cumulative sum of DNLs~\cite{6198279}. Figure~\ref{fig:Hist}b and c show the DNL and INL for V$_{ov}$~=~2~V and 25$^{\circ}$C, respectively. As visible in the histogram, the fine TDCs do not tap their full potential of 32~bins (5~bits). That means, the delay of the DEs in the DLL is too large.
We believe, that the circuit is processed in a process corner, where all transistors work very slowly. Process, voltage and temperature variations can have a big impact on chip functionality. Corner simulations show a delay range (bin width) of 70~ps until 110~ps. To tune these values two control switches (\emph{TDC$\_$cntr<0:1>}) are included in the calibration circuit of the TDC (cf. figure~\ref{fig:circuits}b). But for the slow corner, the additional switched current is insufficient, and only a bin width of about 94~ps is achieved instead of typical 76.5~ps. Furthermore, the distribution of clocks over the matrix can provoke run-time discrepancies. This can be the reason for the different number of fine-TDC bins for the four quadrant TDCs.
	
For the characterization plots in figure~\ref{fig:Hist}, first the DNL standard deviation (cf. $\sigma$DNL in figure~\ref{fig:Hist}b) is determined excluding the last two bins. A 5-$\sigma$ limit on $\sigma$DNL defines the maximal bin number of the fine TDC or the minimal bin width of the last bin, respectively. Second, the DNL and INL are determined for the updated fine-TDC bins.
This procedure was done for all dark-event measurements leading to similar results. Here, the effect of temperature variations is visible in the reached number of bins. The DE delay decreases with increasing temperature. At $-$25$^{\circ}$C, the dynamic range is increased by about two bins (cf. figure~\ref{fig:Hist}d). This effect makes a stable temperature environment desirable, and the differences in bin width have to be considered for the ToA calculations.

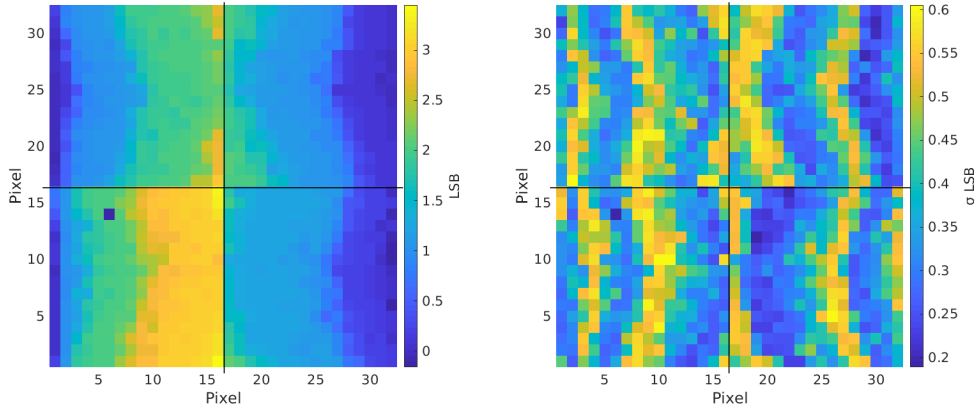
\begin{figure}[h]
	\centering
	\includegraphics[width=1\linewidth]{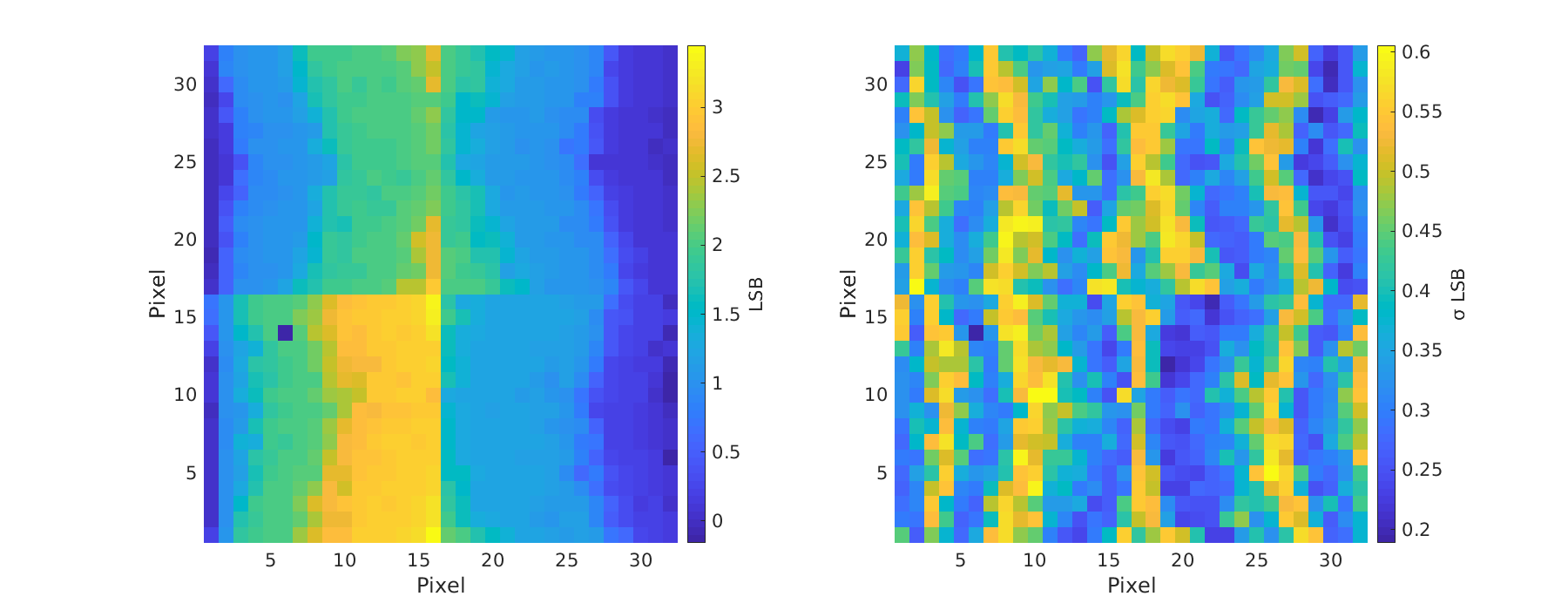}
	\put(-309,16){\line(0,1){140}}	\put(-377,86){\line(1,0){135}}	\put(-120,16){\line(0,1){140}}	\put(-187,86){\line(1,0){135}}
	\caption{Offset map (left), and standard deviation of the offset map (right). Pixel 6/14 (dark blue) was turned off during measurements and its data are neglected.}
	\label{fig:Offset}
\end{figure}

As mentioned in section~\ref{sec:design}, all pixel outputs in a quadrant are connected via wired-ORs triggering the TDC. These metal interconnections come along with parasitic elements implicating propagation delays across the matrix. These delays introduce non-negligible offsets per pixel, which must be considered in timestamp calculations. Figure~\ref{fig:Offset} shows the offset map (left) and a map with its standard deviations (right), measured by using the laser source. For this measurement the pixels are sequentially enabled and illuminated, and the TDC output per pixel is stored. The offset is calculated by subtracting each pixel value by the value of the closest pixel to the corresponding TDC (e.g. pixel 1/9 for quadrant~3), which is expected to be the lowest one. In this case, the offset map illustrates the wired-OR routing to the four TDCs (signal propagating row-wise from middle to the left and right, and then to the middle of the edge of each quadrant). The maximum offset in the area of 1.12~$\times$~1.22~mm$^{2}$ (one quadrant) is (3.45~$\pm$~0.61)~LSB ($\sim$326~ps~$\pm$~86~ps). The minimum offset is $\leq$~0~LSB caused by some jitter. The standard deviation map on the right side of figure~\ref{fig:Offset} highlights the ranges, where the propagation-delay value falls in a LSB change of the TDCs.
	
\subsubsection{Stand-alone TDC}

\begin{figure}[t]
	\centering
	\includegraphics[width=0.98\linewidth,trim=40 40 40 30,clip]{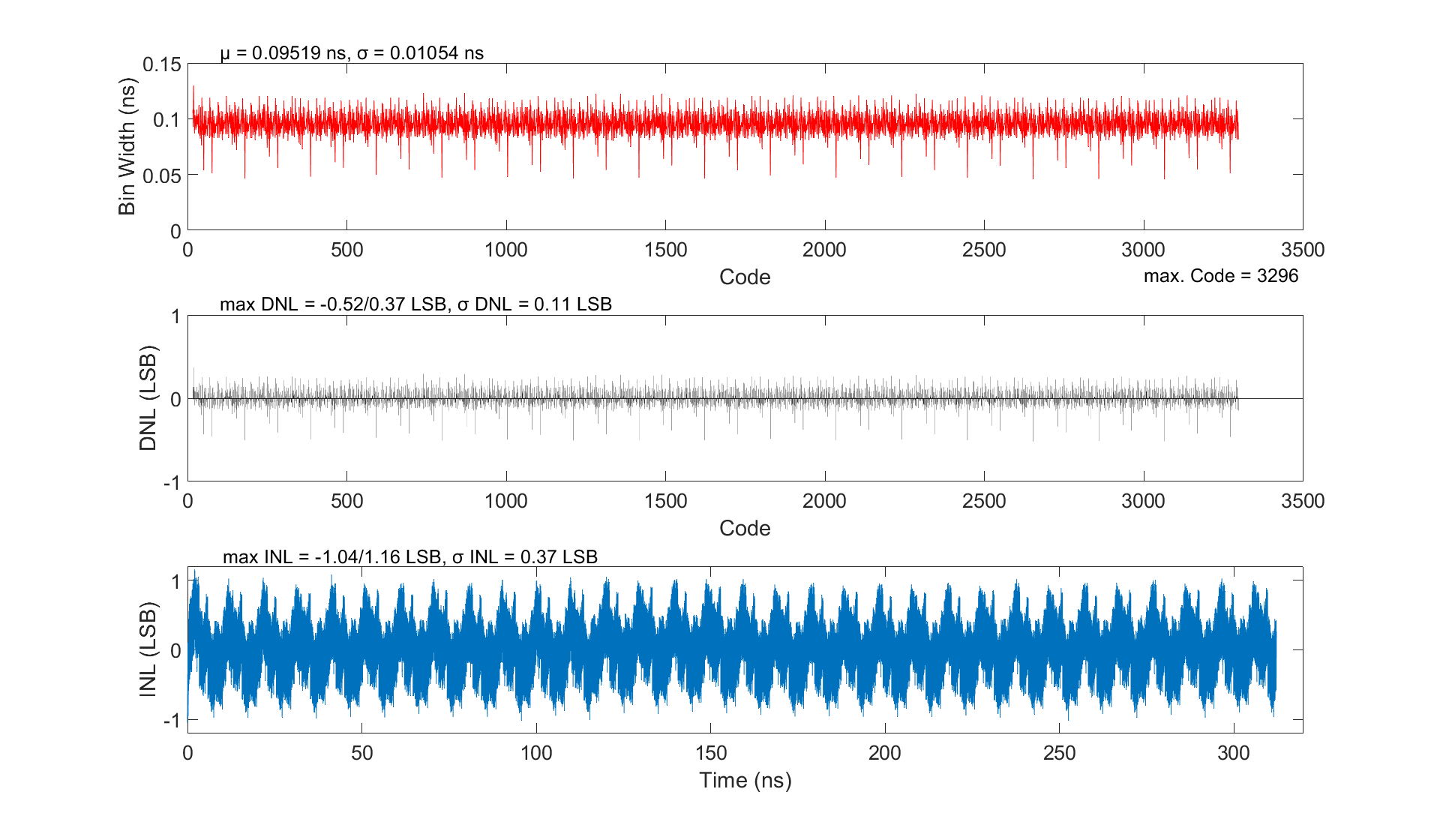}
	\caption{TDC-bin width, DNL and INL (top to bottom) of one sample at 408~MHz reference clock.}
	\label{fig:TDCout}
\end{figure}

\begin{figure}[b]
	\centering
	\includegraphics[width=1\linewidth,trim=20 30 20 30,clip]{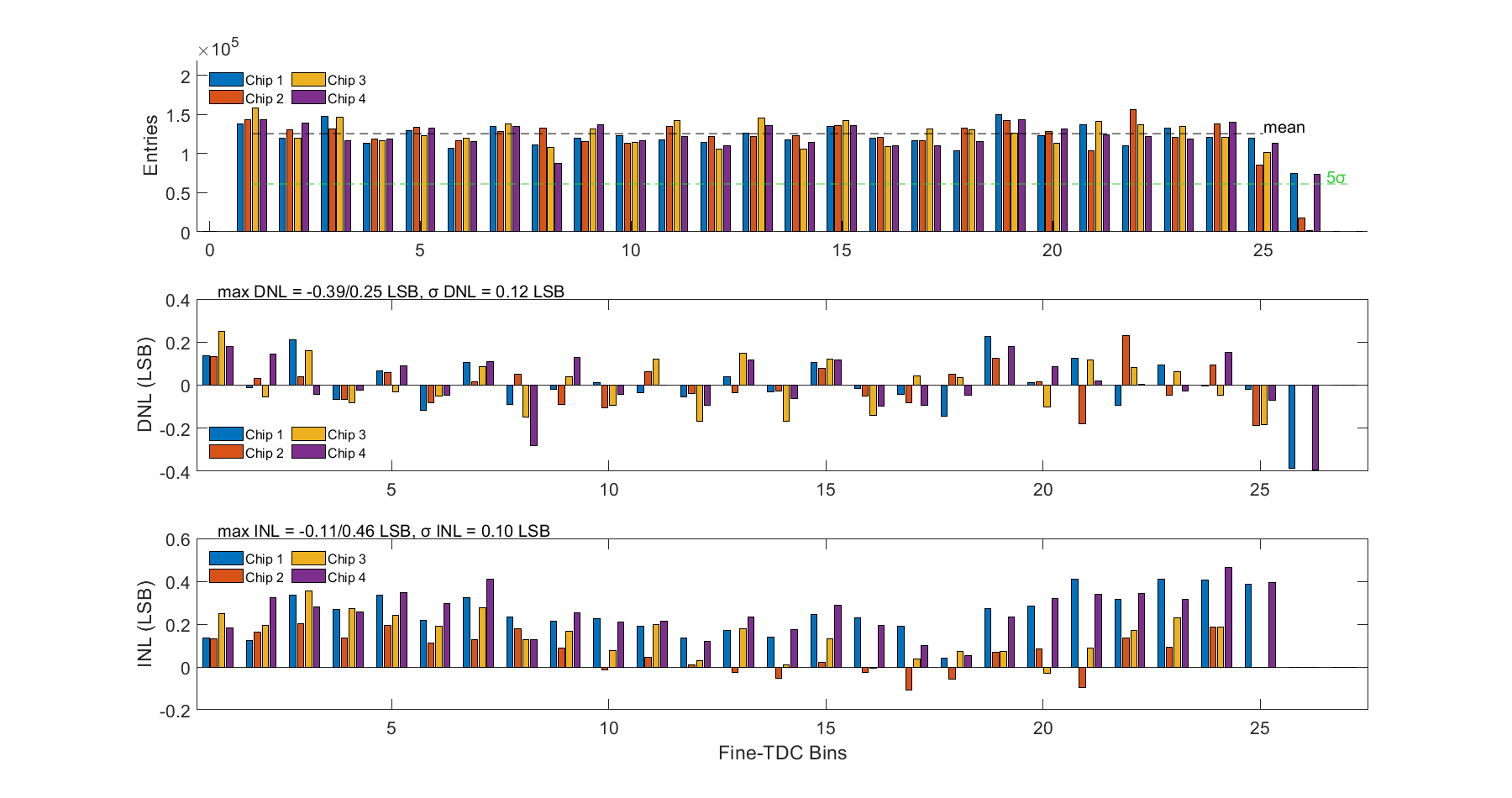}
	\caption{Histogram, DNL and INL (top to bottom) of four sample's fine-TDC data at room temperature.}
	\label{fig:Hist3}
\end{figure}

In contrast to the previous subsection, where random dark events were used to characterize the quadrant fine TDCs, a time-ramp signal was created by driving the trigger input of the stand-alone TDC with step size of 1~ps over the entire dynamic range. As reference clock, the 408-MHz system clock was used. The measured average bin width amounts to $\sim$95~ps. The deviation of each bin width to the average value defines the DNL, and the deviation to an ideal line fitted into the step curve defines the INL. Figure~\ref{fig:TDCout} exemplarily shows the measured TDC characteristics of one sample. Displayed are the bin width and DNL versus code number, and the INL versus time. Additionally, all measured fine-TDC data are plotted into a histogram, and DNL and INL are determined with the same method like for the quadrant fine TDCs. The results for all samples are shown in figure~\ref{fig:Hist3}. The stand-alone TDCs behave very similar to the quadrant TDCs. The reached bit resolution is 11.67~bits. The maximum 12~bit could be reached by decreasing the reference clock to 365~MHz resulting in an average bin width of 86~ps.

\subsubsection{LVDS links}
	
\begin{figure}[h]
		\centering
		\includegraphics[width=0.7\linewidth]{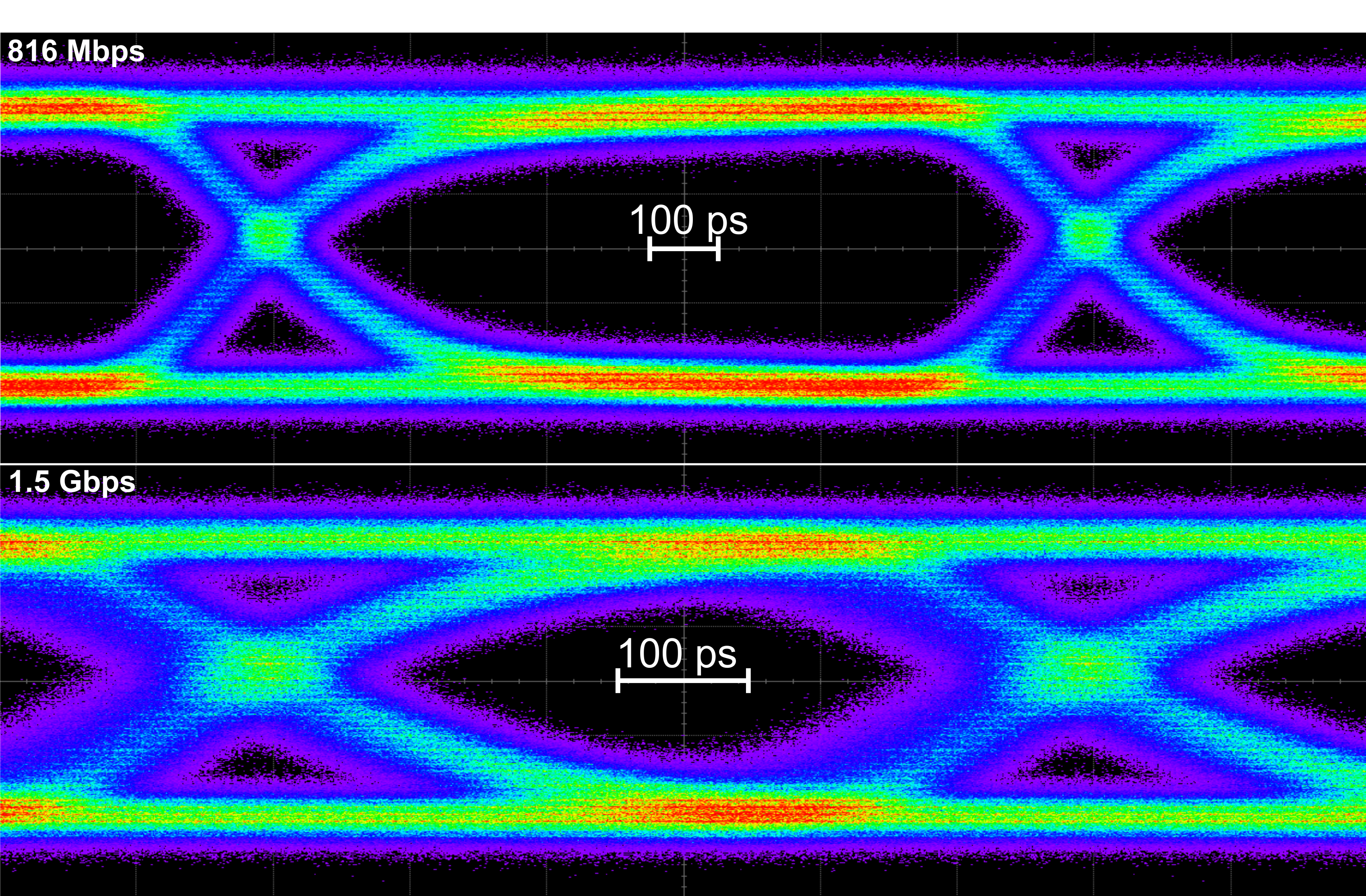}
		\caption{Eye diagrams for the RX-TX chain at 816~Mbps (top) and 1.5~Gbps (bottom).}
		\label{fig:Eyes}
\end{figure}

The test RX and TX (cf. figure~\ref{fig:RXTX}) are connected as input and output for a buffer with high driving strength. The performance of this RX-TX chain was measured via eye diagrams. Figure~\ref{fig:Eyes} shows the eye diagrams at the typical data rate of 816~Mbps (top) and at maximum data rate of 1.5~Gbps. At 816~Mbps, 1.3~Gbps and 1.5~Gbps, a bit-error-rate (BER) of <10$^{-21}$, <10$^{-15}$ and <10$^{-9}$ was achieved, respectively.
	
\subsection{Summary}

The key characteristics are summarized in table~\ref{tab:circuits}, and values obtained in our former design are listed for comparison.
In case of the SPAD characteristics, the DCR per $\mu$m$^{2}$ could be improved by about one order of magnitude. The large spread in the noise behavior indicates that a masking of individual SPAD cells in each pixel could help to improve the DCR without losing all four SPADs per pixel. A higher fill factor could be achieved by using customized SPAD cells.
The simplification of the pixel electronics shared by four SPADs led to a reduction of the area by about 80~$\%$. Waiving the in-pixel hit-counting capability would further reduce the area by 50~$\%$.
The TDC characteristics listed in table~\ref{tab:circuits} are based on the typical system clock of 408~MHz. We obtained very similar characteristics compared to our former design. However, the TDC power and area requirements increased by about 140~$\%$ and 40~$\%$, respectively. This increase is mainly caused by the higher supply voltage, different process node and some modifications in the encoder. Taking the current core area into account, a column-level TDC approach is feasible on the cost of a 16-times higher power consumption and data throughput.
The LVDS links allow for a 60~$\%$ higher speed level than originally targeted (816~Mbps~@~408~MHz). Therewith, the current design permits a maximum frame rate of about 4.8~MHz. In contrast to our former design, the area requirement is increased by 64~$\%$ for TX and RX. The reason is mainly the different process with other resistor sizes. The conservation of 12-bit resolution entails the adjustment of the ToA-bin width.

\newpage	
\begin{table}[h]
	\centering
	\caption{Key characteristics of the SPAD, pixel electronics, TDC and LVDS links, determined at room temperature.\label{tab:circuits}}
	\smallskip
	\begin{minipage}{130mm}

	\begin{tabular}{|c r|c|c|}
		\hline
		\multicolumn{2}{|c|}{Parameter}&This work&\cite{8824395}\\
		\hline\hline
		\multicolumn{2}{|c|}{CMOS node (nm)}&150&130\\
		\hline
		\multicolumn{4}{|c|}{SPAD}\\
		\hline
		\multicolumn{2}{|c|}{Pixel pitch (µm)}&70&50\\
		\hline
		\multicolumn{2}{|c|}{Configuration}&32~$\times$~32&16~$\times$~16\\
		\hline
		\multicolumn{2}{|c|}{Fill factor ($\%$)}&30&90\\
		\hline
		\multicolumn{2}{|c|}{mean DCR (Hz/($\mu$m$^{2}$) @ V$_{ov}$~=~1~V}&8.7 & 80 \\
		\hline
		\multicolumn{4}{|c|}{Pixel electronics}\\
		\hline
		\multicolumn{2}{|c|}{Area ($\mu$m~$\times~\mu$m)}&70~$\times$~5 + 3~$\times$~17\footnote{cf. figure~\ref{fig:chip}b}&40~$\times$~45\\
		\hline
		\multicolumn{2}{|c|}{Power ($\mu$W)}& 10 & 25\\
		\hline
		\multicolumn{4}{|c|}{TDC}\\
		\hline
		\multicolumn{2}{|c|}{Resolution (bit)}&11.67&12\\
		\hline
		\multicolumn{2}{|c|}{Precision (ps)}&95.8~$\pm~$13.65&77.19~$\pm$~7.53 \\
		\hline
		\multicolumn{2}{|c|}{max.~DNL (LSB)}&$-$0.74~/~0.35&$-$0.46~/~0.64\\
		\hline
		\multicolumn{2}{|c|}{max.~INL (LSB)}&$-$1.43~/~1.39&$-$1.33~/~0.93\\
		\hline
		\multicolumn{2}{|c|}{Power (mW)}&11&4.6\\
		\hline
		\multicolumn{2}{|c|}{Area ($\mu$m~$\times~\mu$m)}&78~$\times$~157&55~$\times$~160\\
		\hline
		\multicolumn{4}{|c|}{Links (RX-TX chain)}\\
		\hline
		&@BER=10$^{-15}$&1.3&1.2\\
		\raisebox{1.5ex}[-1.5ex]{max. data throughput (Gbps)}&@BER=10$^{-9~}$&1.5&1.6\\
		\hline
		Power (mW)&RX~/~TX&3~/~37&8~/~48\\
		\hline
		  &RX&39~$\times$~38&30~$\times$~30\\
		\raisebox{1.5ex}[-1.5ex]{Area ($\mu$m~$\times$~$\mu$m)}&TX&73~$\times$~94&60~$\times$~70\\
		\hline

	\end{tabular}
	\end{minipage}
\end{table}

		\section{Conclusions} \label{sec:conclude}
	We presented a monolithic 32~$\times$~32 dSiPM-pixel matrix IC with 70~$\mu$m pitch and 30~$\%$ fill factor designed and fabricated in LF’s 150-nm CMOS technology. It enables full hit-map readout and provides sub-100~ps time stamping for each quadrant. Main focus has been taken on dark-event measurements in a climate chamber to show the dependency of DCR on temperature and overvoltage. Also the event-related timestamps for different temperatures have been analyzed. With aid of a pulsed laser, propagation delays across the matrix have been determined. Furthermore, the TDC resolution, DNL and INL as well as data transmission speed limits have been identified and compared with a previous prototype designed for a hybrid concept.

\acknowledgments

The authors would like to thank A.~Venzmer, E.~Wüstenhagen and D.~Gorski for test-board and mechanical case design, test setup, as well as chip assembly. We are grateful to C.~Reckleben and S.~Lachnit for fruitful discussions and manuscript reading.

	
\end{document}